\documentclass[11pt]{article}
\usepackage{amsmath,amssymb,amsthm,amsxtra,overpic,bbm,bm,epsfig,ulem,color,multirow}
\usepackage{braket}

\begin{document}

\begin{center}
{\Large Purely flavored leptogenesis from a sudden mass gain \\
 of right-handed neutrinos}
\end{center}

\vspace{0.05cm}

\begin{center}
{\bf Zhen-hua Zhao\footnote{zhaozhenhua@lnnu.edu.cn}, Jing Zhang, Xiang-Yi Wu} \\
{ $^1$ Department of Physics, Liaoning Normal University, Dalian 116029, China \\
$^2$ Center for Theoretical and Experimental High Energy Physics, \\ Liaoning Normal University, Dalian 116029, China }
\end{center}

\vspace{0.2cm}

\begin{abstract}
In this paper, we would like to point out that in the scenario that the right-handed neutrinos suddenly gain some masses much larger than the temperature of the Universe at that time so that the washout effects for the lepton asymmetry generated from their decays can be neglected safely, the purely flavored leptogenesis scenario (in which the total CP asymmetries for the decays of the right-handed neutrinos are vanishing and the successful leptogenesis is realized by virtue of the flavor non-universality of the washout effects) cannot work in the usual way any more. For this problem, we put forward that the flavor non-universality of the conversion efficiencies from the flavored lepton asymmetries to the baryon asymmetry via the sphaleron processes may play a crucial role.
And we will study if the requisite baryon asymmetry can be successfully reproduced from such a mechanism in the scenarios that the right-handed neutrino masses are hierarchical and nearly degenerate, respectively. A detailed study shows that this mechanism can be viable in both these two scenarios.
\end{abstract}

\newpage

\section{Introduction}

As we know, the phenomena of neutrino oscillations indicate that neutrinos are massive and their flavor eigenstates $\nu^{}_\alpha$ (for $\alpha =e, \mu, \tau$) are certain superpositions of the mass eigenstates $\nu^{}_i$ (for $i =1, 2, 3$) with definite masses $m^{}_i$: $\nu^{}_\alpha = \sum^{}_i U^{}_{\alpha i} \nu^{}_i$ with $U^{}_{\alpha i}$ being the $\alpha i$ element of the $3 \times 3$  neutrino mixing matrix $U$ \cite{xing}. In the standard parametrization, $U$ is expressed in terms of three mixing angles $\theta^{}_{ij}$ (for $ij=12, 13, 23$), one Dirac CP phase $\delta$ and two Majorana CP phases $\rho$ and $\sigma$ as
\begin{eqnarray}
U  = \left( \begin{matrix}
c^{}_{12} c^{}_{13} & s^{}_{12} c^{}_{13} & s^{}_{13} e^{-{\rm i} \delta} \cr
-s^{}_{12} c^{}_{23} - c^{}_{12} s^{}_{23} s^{}_{13} e^{{\rm i} \delta}
& c^{}_{12} c^{}_{23} - s^{}_{12} s^{}_{23} s^{}_{13} e^{{\rm i} \delta}  & s^{}_{23} c^{}_{13} \cr
s^{}_{12} s^{}_{23} - c^{}_{12} c^{}_{23} s^{}_{13} e^{{\rm i} \delta}
& -c^{}_{12} s^{}_{23} - s^{}_{12} c^{}_{23} s^{}_{13} e^{{\rm i} \delta} & c^{}_{23}c^{}_{13}
\end{matrix} \right) \left( \begin{matrix}
e^{{\rm i}\rho} &  & \cr
& e^{{\rm i}\sigma}  & \cr
&  & 1
\end{matrix} \right) \;,
\label{1}
\end{eqnarray}
where the abbreviations $c^{}_{ij} = \cos \theta^{}_{ij}$ and $s^{}_{ij} = \sin \theta^{}_{ij}$ have been employed.

Thanks to the various neutrino oscillation experiments, the neutrino mixing angles and neutrino mass squared differences $\Delta m^2_{ij} \equiv m^2_i - m^2_j$ have been measured to a good degree of accuracy, and there is also a preliminary result for $\delta$ (but with a large uncertainty). Several research groups have performed global analyses of the accumulated neutrino oscillation data to extract the values of these parameters \cite{global,global2}. For definiteness, we will use the results in Ref.~\cite{global} (reproduced in Table~1 here) as reference values in the following numerical calculations: the best-fit values of the neutrino mixing angles and neutrino mass squared differences will be taken as typical inputs, while $\delta$ will be treated as a free parameter in consideration of its large uncertainty.
Note that the sign of $\Delta m^2_{31}$ remains undetermined, thereby allowing for two possible neutrino mass orderings: the normal ordering (NO) $m^{}_1 < m^{}_2 < m^{}_3$ and inverted ordering (IO) $m^{}_3 < m^{}_1 < m^{}_2$. However, neutrino oscillations are completely insensitive to the absolute neutrino mass scale and the Majorana CP phases. Their values can only be inferred from certain non-oscillatory experiments such as the neutrinoless double beta decay experiments \cite{0nbb}. But so far there has not been any lower bound on the lightest neutrino mass, nor any constraint on the Majorana CP phases.

On the other hand, one of the most popular and natural ways of generating the tiny neutrino masses is the type-I seesaw model in which two or three heavy right-handed neutrinos $N^{}_I$ ($I=1, 2, 3$) are introduced into the Standard Model (SM) \cite{seesaw}. First of all, $N^{}_I$ can constitute the Yukawa coupling operators together with the left-handed neutrinos $\nu^{}_\alpha$ (which reside in the lepton doublets $L^{}_\alpha$) and the Higgs doublet $H$: $(Y^{}_{\nu})^{}_{\alpha I} \overline {L^{}_\alpha} H N^{}_I $ with $(Y^{}_{\nu})^{}_{\alpha I}$ being the $\alpha I$ element of the Yukawa coupling matrix $Y^{}_{\nu}$. These operators will generate the Dirac neutrino masses $(M^{}_{\rm D})^{}_{\alpha I}= (Y^{}_{\nu})^{}_{\alpha I} v$ [here $(M^{}_{\rm D})^{}_{\alpha I}$ is the $\alpha I$ element of the Dirac neutrino mass matrix $M^{}_{\rm D}$] after the neutral component of $H$ acquires the nonzero vacuum expectation value (VEV) $v = 174$ GeV. Furthermore, $N^{}_I$ themselves can also have the Majorana mass terms $\overline{N^c_I} (M^{}_{\rm R})^{}_{IJ} N^{}_J$ [here $(M^{}_{\rm R})^{}_{IJ}$ is the $IJ$ element of the right-handed neutrino mass matrix $M^{}_{\rm R}$].
Then, under the seesaw condition $M^{}_{\rm R} \gg M^{}_{\rm D}$, one will obtain an effective Majorana mass matrix for the three light neutrinos as
\begin{eqnarray}
M^{}_{\nu} = - M^{}_{\rm D} M^{-1}_{\rm R} M^{T}_{\rm D} \;,
\label{2}
\end{eqnarray}
by integrating the right-handed neutrinos out. Thanks to such a formula, the smallness of neutrino masses can be naturally explained by the heaviness of right-handed neutrinos. Throughout this paper, without loss of generality, we will work in the basis of $M^{}_{\rm R}$ being diagonal as $D^{}_{\rm R} = {\rm diag}(M^{}_1, M^{}_2, M^{}_3)$ with $M^{}_I$ being the mass of $N^{}_I$ and $M^{}_1 < M^{}_2 < M^{}_3$.

\begin{table}\centering
  \begin{footnotesize}
    \begin{tabular}{c|cc|cc}
     \hline\hline
      & \multicolumn{2}{c|}{Normal Ordering}
      & \multicolumn{2}{c}{Inverted Ordering }
      \\
      \cline{2-5}
      & bf $\pm 1\sigma$ & $3\sigma$ range
      & bf $\pm 1\sigma$ & $3\sigma$ range
      \\
      \cline{1-5}
      \rule{0pt}{4mm}\ignorespaces
       $\sin^2\theta^{}_{12}$
      & $0.303_{-0.012}^{+0.012}$ & $0.270 \to 0.341$
      & $0.303_{-0.012}^{+0.012}$ & $0.270 \to 0.341$
      \\[1mm]
       $\sin^2\theta^{}_{23}$
      & $0.451_{-0.016}^{+0.019}$ & $0.408 \to 0.603$
      & $0.569_{-0.021}^{+0.016}$ & $0.412 \to 0.613$
      \\[1mm]
       $\sin^2\theta^{}_{13}$
      & $0.02225_{-0.00059}^{+0.00056}$ & $0.02052 \to 0.02398$
      & $0.02223_{-0.00058}^{+0.00058}$ & $0.02048 \to 0.02416$
      \\[1mm]
       $\delta/\pi$
      & $1.29_{-0.14}^{+0.20}$ & $0.80 \to 1.94$
      & $1.53_{-0.16}^{+0.12}$ & $1.08\to 1.91$
      \\[3mm]
       $\Delta m^2_{21}/(10^{-5}~{\rm eV}^2)$
      & $7.41_{-0.20}^{+0.21}$ & $6.82 \to 8.03$
      & $7.41_{-0.20}^{+0.21}$ & $6.82 \to 8.03$
      \\[3mm]
       $|\Delta m^2_{31}|/(10^{-3}~{\rm eV}^2)$
      & $2.507_{-0.027}^{+0.026}$ & $2.427 \to 2.590$
      & $2.412_{-0.025}^{+0.028}$ & $2.332 \to 2.496$
      \\[2mm]
      \hline\hline
    \end{tabular}
  \end{footnotesize}
  \caption{The best-fit values, 1$\sigma$ errors and 3$\sigma$ ranges of six neutrino
oscillation parameters extracted from a global analysis of the existing
neutrino oscillation data \cite{global}. }
\end{table}

Remarkably, the seesaw model also provides an attractive explanation (which is known as the leptogenesis mechanism \cite{leptogenesis, Lreview}) for the baryon-antibaryon asymmetry of the Universe \cite{planck}
\begin{eqnarray}
Y^{}_{\rm B} \equiv \frac{n^{}_{\rm B}-n^{}_{\rm \bar B}}{s} \simeq (8.69 \pm 0.04) \times 10^{-11}  \;,
\label{3}
\end{eqnarray}
where $n^{}_{\rm B}$ ($n^{}_{\rm \bar B}$) denotes the baryon (antibaryon) number density and $s$ the entropy density. The leptogenesis mechanism works in a way as follows: a lepton-antilepton asymmetry is first generated from the out-of-equilibrium and CP-violating decays of the right-handed neutrinos and then partly converted into the baryon-antibaryon asymmetry via the sphaleron processes: $Y^{}_{\rm B} = c Y^{}_{\rm L}$ with $c$ being the conversion efficiency from the lepton asymmetry to the baryon asymmetry. At the leading order, $c$ takes a value about $-1/3$.

As is known, according to the temperature where leptogenesis takes place, there are the following three distinct leptogenesis regimes  \cite{flavor}.
(1) Unflavored regime: in the temperature range above $10^{12}$ GeV where the charged lepton Yukawa $y^{}_\alpha$ interactions have not yet entered thermal equilibrium, three lepton flavors are indistinguishable from one another so that they should be treated in a universal way. In this regime, the final baryon asymmetry from the right-handed neutrino $N^{}_I$ is given by
\begin{eqnarray}
Y^{}_{\rm B} = c r \varepsilon^{}_I \kappa(\widetilde m^{}_I)  \;,
\label{4}
\end{eqnarray}
where $r \simeq 4 \times 10^{-3}$ measures the ratio of the equilibrium number density of $N^{}_I$ to the entropy density. And $\varepsilon^{}_I$ is the total CP asymmetry between the decay rates of $N^{}_I \to L^{}_\alpha + H$ and their CP-conjugate processes $N^{}_I \to \overline{L}^{}_\alpha + \overline{H}$.
It is a sum of the flavored CP asymmetries $\varepsilon^{}_{I \alpha}$ (i.e., $\varepsilon^{}_I = \sum^{}_\alpha \varepsilon^{}_{I \alpha}$).
Finally, $\kappa(\widetilde m^{}_I) \leq 1$ is the efficiency factor (i.e., the survival probability of the lepton asymmetry generated from the decays of $N^{}_I$) which takes account of the washout effects due to the inverse decays of $N^{}_I$ and various lepton-number-violating scattering processes. Its concrete value depends on the washout mass parameter
\begin{eqnarray}
\widetilde m^{}_I = \sum^{}_\alpha \widetilde m^{}_{I \alpha} = \sum^{}_\alpha  \frac{|(M^{}_{\rm D})^{}_{\alpha I}|^2}{M^{}_I} \;,
\label{5}
\end{eqnarray}
and can be numerically calculated by solving the relevant Boltzmann equations \cite{Lreview}.
(2) Two-flavor regime: in the temperature range $10^{9}$---$10^{12}$ GeV where the $y^{}_\tau$-related interactions have entered thermal equilibrium, the $\tau$ flavor is distinguishable from the other two flavors which remain indistinguishable from each other so that there are effectively two flavors (i.e., the $\tau$ flavor and a coherent superposition of the $e$ and  $\mu$ flavors). In this regime, the final baryon asymmetry from $N^{}_I$ is given by
\begin{eqnarray}
Y^{}_{\rm B}
=  c r \left[ \varepsilon^{}_{I \gamma} \kappa \left(\frac{417}{589} \widetilde m^{}_{I \gamma} \right) + \varepsilon^{}_{I \tau} \kappa \left(\frac{390}{589} \widetilde m^{}_{I \tau} \right) \right]
 \;,
\label{6}
\end{eqnarray}
with $\varepsilon^{}_{I \gamma} = \varepsilon^{}_{I e} + \varepsilon^{}_{I \mu}$ and $\widetilde m^{}_{I \gamma} = \widetilde m^{}_{I e} + \widetilde m^{}_{I \mu}$. Here the factors 417/589 and 390/589 correspond to the diagonal entries of the $A$ matrix and quantifies the effects of flavor in the washout processes when changing from the lepton asymmetries to the baryon asymmetry (for more details, see Ref.~\cite{flavor}).
(3) Three-flavor regime: in the temperature range below $10^{9}$ GeV where the $y^{}_\mu$-related interactions have also entered thermal equilibrium, all the three lepton flavors are distinguishable from one another so that they should be treated separately. In this regime, the final baryon asymmetry from $N^{}_I$ is given by
\begin{eqnarray}
Y^{}_{\rm B} = c r \left[ \varepsilon^{}_{I e} \kappa \left(\frac{453}{537} \widetilde m^{}_{I e} \right) + \varepsilon^{}_{I \mu} \kappa \left(\frac{344}{537} \widetilde m^{}_{I \mu} \right) + \varepsilon^{}_{I \tau} \kappa \left(\frac{344}{537} \widetilde m^{}_{I\tau} \right) \right] \; .
\label{7}
\end{eqnarray}

In the literature, a physically interesting and extensively studied possibility \cite{flavored} is when the leptogenesis mechanism is completely realized through the flavor effects: in this scenario, the total CP asymmetry $\varepsilon^{}_I= \sum^{}_\alpha \varepsilon^{}_{I \alpha}$ is vanishing while the flavored CP asymmetries $\varepsilon^{}_{I \alpha}$ are individually non-vanishing (which can be naturally realized in flavour models with residual CP symmetries \cite{rCP}). Consequently, the leptogenesis mechanism would fail to work in the unflavored regime [i.e., $Y^{}_{\rm B} =0$ as can be seen from Eq.~(\ref{4})] \footnote{ It is interesting to note that in the literature there are the following two exceptional scenarios: a non-vanishing $Y^{}_{\rm B}$ is still possible in the unflavored regime even if $\sum^{}_\alpha \varepsilon^{}_{I \alpha} =0$ holds when the quantum density matrix formalism is taken into account \cite{matrix} or after the inclusion of quantum corrections to the
Casas-Ibarra parameterisation \cite{RGE}. }. On the other hand, in the two-flavor or three-flavor regime, since the CP asymmetries in different flavors are subject to different washout effects [as can be seen from Eqs.~(\ref{6}, \ref{7})], the leptogenesis mechanism still can work in spite of $\sum^{}_\alpha \varepsilon^{}_{I \alpha} =0$. Just for this reason, this scenario will be referred to as purely flavored leptogenesis (i.e., only when the flavor effects come into play can the leptogenesis mechanism work).

Another physically interesting possibility is that the right-handed neutrino are initially protected to be massless by some symmetry (e.g., the B$-$L symmetry \cite{BL}) and then suddenly become massive at a critique temperature $T^{}_{\rm c}$ via their couplings with a scalar field which spontaneously breaks this symmetry by acquiring a non-vanishing VEV. As noted in Ref.~\cite{massgain}, if the gained masses of the right-handed neutrinos are much larger than the temperature of the Universe at that time (i.e., $T^{}_{\rm c}$), they will fall into a state featuring strong departure from thermal equilibrium and consequently decay and generate a nonzero lepton asymmetry very rapidly. And the inverse decays will be immediately Boltzmann suppressed so that the washout effects for the generated lepton asymmetry can be neglected safely (equivalent to taking the efficiency factors $\kappa'$s to be unity). To be specific, we will assume the right-handed neutrino masses to be larger than $T^{}_{\rm c}$ by at least 20 times so that the washout effects will be Boltzmann suppressed by at least the order ${\cal{O}}(e^{-M^{}_I/T^{}_{\rm c}}) \sim {\cal{O}}(10^{-9})$ and thus can be neglected safely.

In this paper, we would like to point out that in the scenario that the right-handed neutrinos suddenly gain some masses much larger than the temperature of the Universe at that time so that the washout effects for the generated lepton asymmetry can be neglected safely, the purely flavored leptogenesis (which crucially relies on the flavor non-universality of the washout effects) cannot work in the usual way any more. For this problem, we put forward that the flavor non-universality of the conversion efficiencies from the lepton asymmetries to the baryon asymmetry via the sphaleron processes may play a crucial role: the complete expression for the relation between the baryon and lepton asymmetries takes a form as (see Ref.~\cite{sphaleron} for more details)
\begin{eqnarray}
Y^{}_{\rm B} & = & -4 \frac{77 T^2+54v^2}{869 T^2+666 v^2} \sum^{}_\alpha Y^{}_{{\rm L}\alpha}  \nonumber \\
&  & -
\left(\frac{11 v^2}{2\pi^2 T^2}
\frac{47 T^2+36 v^2}{869 T^2+666 v^2}
+ \frac{1}{16\pi^2} \frac{1034 T^2+810 v^2}{869 T^2+666 v ^2} \right)
\sum^{}_\alpha y_\alpha^2 Y^{}_{{\rm L}\alpha}  \;,
\label{8}
\end{eqnarray}
where $T \simeq 135$ GeV is the decoupling temperature of the sphaleron process \cite{sphaleron2}, $v=174$ GeV the Higgs VEV, $Y^{}_{{\rm L}\alpha}$ the flavored lepton asymmetries and $y^{}_\alpha$ the Yukawa coupling coefficients of the charged leptons. One can see that in the first term the conversion efficiency (which has a value about $-1/3$)  from the lepton asymmetries to the baryon asymmetry is flavor universal and it is just the factor that is commonly used as the conversion efficiency from the lepton asymmetry to the baryon asymmetry. On the other hand, in the second term the conversion efficiencies from the lepton asymmetries to the baryon asymmetry are flavor dependent (controlled by $y^2_\alpha$). Although the conversion coefficients in the second term are highly suppressed by $y^2_\alpha$ (concretely, $y^2_e \simeq 8.3 \times 10^{-12}$, $y^2_\mu \simeq 3.7 \times 10^{-7}$ and $y^2_\tau \simeq 1.0 \times 10^{-4}$), they may play a key role in the scenario that the total lepton asymmetry is vanishing: $Y^{}_{\rm L} =\sum^{}_\alpha Y^{}_{{\rm L}\alpha}=0$ [which would render the first term in Eq.~(\ref{8}) to be vanishing].
Thanks to such an effect, in the scenario considered in this paper which just realizes $\sum^{}_\alpha Y^{}_{{\rm L}\alpha}=0$ (as a joint result of $\sum^{}_\alpha \varepsilon^{}_{I \alpha} =0$ and the absence of the washout effects), a non-vanishing $Y^{}_{\rm B}$ is still possible. To be specific, in the scenario considered in this paper, Eq.~(\ref{8}) gives
\begin{eqnarray}
Y^{}_{\rm B} \simeq - 0.06 y^2_\tau Y^{}_{{\rm L} \tau} \simeq - 2.4 \times 10^{-8} \sum^{}_I \varepsilon^{}_{I \tau}  \;.
\label{9}
\end{eqnarray}
This tells us that, provided that $ \varepsilon^{}_{I \tau}$ is sizable enough (i.e., $-3.6 \times 10^{-3}$), the observed value of $Y^{}_{\rm B}$ can be successfully reproduced. In the following two sections, we will study if the requisite baryon asymmetry can be successfully reproduced from such a mechanism in the scenarios that the right-handed neutrino masses are hierarchical and nearly degenerate, respectively.

\section{Study for hierarchical right-handed neutrino masses}

In this section, we perform the study in the scenario that the right-handed neutrino masses are hierarchical, in which scenario the flavored CP asymmetries are explicitly expressed as
\begin{eqnarray}
&& \varepsilon^{}_{I \alpha} = \frac{1}{8\pi (M^\dagger_{\rm D}
M^{}_{\rm D})^{}_{II} v^2} \sum^{}_{J \neq I} \left\{ {\rm Im}\left[(M^*_{\rm D})^{}_{\alpha I} (M^{}_{\rm D})^{}_{\alpha J}
(M^\dagger_{\rm D} M^{}_{\rm D})^{}_{IJ}\right] {\cal F} \left( \frac{M^2_J}{M^2_I} \right) \right. \nonumber \\
&& \hspace{1.cm}
+ \left. {\rm Im}\left[(M^*_{\rm D})^{}_{\alpha I} (M^{}_{\rm D})^{}_{\alpha J} (M^\dagger_{\rm D} M^{}_{\rm D})^*_{IJ}\right] {\cal G}  \left( \frac{M^2_J}{M^2_I} \right) \right\} \; ,
\label{2.1}
\end{eqnarray}
with ${\cal F}(x) = \sqrt{x} \{(2-x)/(1-x)+ (1+x) \ln [x/(1+x)] \}$ and ${\cal G}(x) = 1/(1-x)$. And the total CP asymmetries are obtained as
\begin{eqnarray}
&& \varepsilon^{}_{I} = \frac{1}{8\pi (M^\dagger_{\rm D}
M^{}_{\rm D})^{}_{II} v^2} \sum^{}_{J \neq I}  {\rm Im} \left[ (M^\dagger_{\rm D} M^{}_{\rm D})^{2}_{IJ}\right] {\cal F} \left( \frac{M^2_J}{M^2_I} \right)  \;.
\label{2.2}
\end{eqnarray}

In order to facilitate the study, we will employ the popular and convenient Casas-Ibarra parametrization of $M^{}_{\rm D}$ \cite{CI}:
\begin{eqnarray}
M^{}_{\rm D} = {\rm i} U \sqrt{D^{}_\nu} O \sqrt{D^{}_{\rm R}}  \;,
\label{2.3}
\end{eqnarray}
with $\sqrt{D^{}_\nu} = {\rm diag}(\sqrt{m^{}_1}, \sqrt{m^{}_2}, \sqrt{m^{}_3})$ and $\sqrt{D^{}_{\rm R}} = {\rm diag}(\sqrt{M^{}_1}, \sqrt{M^{}_2}, \sqrt{M^{}_3})$. Here $O$ is a complex orthogonal matrix satisfying $O^T O =I$. With the help of such a parametrization, one can directly verify that the condition $\sum^{}_\alpha \varepsilon^{}_{I \alpha} =0$ can be realized in the case that the elements of $O$ are either real or purely imaginary. To be concrete, there are the following four forms of $O$ that can fulfill this condition:
\begin{eqnarray}
O = O^{}_x O^{}_y O^{}_z  \;, \hspace{1cm}
O = O^{}_x O^{\prime}_y O^{\prime}_z \;, \hspace{1cm}
O = O^{\prime}_x O^{}_y O^{\prime}_z \;, \hspace{1cm}
O = O^{\prime}_x O^{\prime}_y O^{}_z \;,
\label{2.4}
\end{eqnarray}
with
\begin{eqnarray}
O^{}_x  = \left( \begin{matrix}
1 & 0 & 0 \cr
0 & \cos x  & \xi \sin x \cr
0 & -\sin x &  \xi \cos x
\end{matrix} \right) \;, \hspace{1cm}
O^\prime_x  = \left( \begin{matrix}
1 & 0 & 0 \cr
0 & \cosh x  & {\rm i}  \xi \sinh x \cr
0 & - {\rm i} \sinh x &  \xi \cosh x
\end{matrix} \right) \;, \nonumber \\
O^{}_y = \left( \begin{matrix}
\cos y & 0 &  \xi \sin y \cr
0 & 1  & 0 \cr
-\sin y & 0 &  \xi \cos y
\end{matrix} \right) \;, \hspace{1cm}
O^\prime_y = \left( \begin{matrix}
\cosh y & 0 & {\rm i}  \xi \sinh y \cr
0 & 1  & 0 \cr
-{\rm i} \sinh y & 0 &  \xi \cosh y
\end{matrix} \right) \;, \nonumber \\
O^{}_z = \left( \begin{matrix}
\cos z &  \xi \sin z & 0 \cr
-\sin z &  \xi \cos z  & 0 \cr
0 & 0 & 1
\end{matrix} \right) \;, \hspace{1cm}
O^\prime_z = \left( \begin{matrix}
\cosh z & {\rm i}  \xi \sinh z & 0 \cr
-{\rm i} \sinh z &  \xi \cosh z  & 0 \cr
0 & 0 & 1
\end{matrix} \right) \;,
\label{2.5}
\end{eqnarray}
where $x$, $y$ and $z$ are real parameters and $\xi = \pm 1$ [corresponding to ${\rm det}(O) = \pm1$].
As in Ref.~\cite{xi}, we choose to work with $\xi = 1$ but extend the range of the Majorana CP
phases $\rho$ and $\sigma$ from $[0, \pi]$ to $[0, 2\pi]$ in our numerical calculations. In this way, the same full sets of CI and Yukawa matrices are considered.

For simplicity and clarity, we will study if the observed value of $Y^{}_{\rm B}$ can be successfully reproduced in the cases that only one of $x$, $y$ and $z$ is non-vanishing (for model realizations of such cases, see Refs.~\cite{rCP, zhao}). Before proceeding, it should be noted that in the usual seesaw mechanism the final baryon asymmetry mainly comes from the lightest right-handed neutrino, since its related processes will effectively washout the lepton asymmetries generated from the heavier right-handed neutrinos. But in the scenario considered in this paper, due to the absence of the washout effects, the contributions of the heavier right-handed neutrinos to the final baryon asymmetry may become significant and will also be taken into account.

\begin{figure*}
\centering
\includegraphics[width=6.5in]{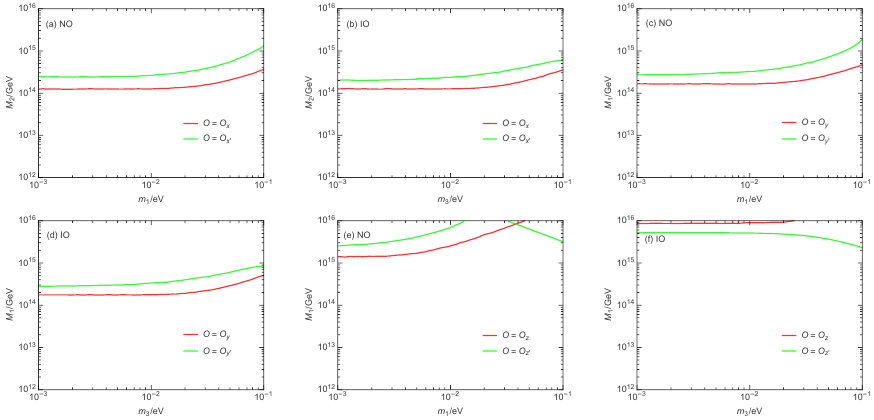}
\caption{ For the cases of $O=O^{}_x$, $O^{}_y$, $O^{}_z$, $O^\prime_x$, $O^\prime_y$ and $O^\prime_z$, the minimal values of the right-handed neutrino masses needed to reproduce the observed value of $Y^{}_{\rm B}$ as functions of the lightest neutrino mass $m^{}_1$ and $m^{}_3$ in the NO and IO cases, respectively.  }
\label{fig1}
\end{figure*}

Let us first perform the study for the case of $O=O^{}_x$. In this case, $N^{}_1$ will decouple from leptogenesis and its only role is to be responsible for the generation of $m^{}_1$. This point can be easily understood by noting that such a form of $O$ will lead us to have $\varepsilon^{}_{1 \alpha} = 0$ as a result of $(M^\dagger_{\rm D} M^{}_{\rm D})^{}_{1J} =0$ for $J\neq 1$. For this reason, one just needs to consider the contributions of $N^{}_2$ and $N^{}_3$ to the final baryon asymmetry. By substituting Eq.~(\ref{2.3}) into Eq.~(\ref{2.1}), one arrives at
\begin{eqnarray}
&& \varepsilon^{}_{2 \tau} = \frac{M^{}_3 \sqrt{m^{}_2 m^{}_3 } (m^{}_2 - m^{}_3) \sin 2x  }{16\pi v^2 (m^{}_2 \cos^2 x+ m^{}_3 \sin^2 x) }  \left[ {\cal F} \left( \frac{M^2_3}{M^2_2} \right) + {\cal G}  \left( \frac{M^2_3}{M^2_2} \right) \right] \Delta^{}_x \;, \nonumber \\
&& \varepsilon^{}_{3 \tau} = \frac{M^{}_2 \sqrt{m^{}_2 m^{}_3 } (m^{}_3 - m^{}_2) \sin 2x  }{16\pi v^2 (m^{}_2 \sin^2 x+ m^{}_3 \cos^2 x) }  \left[ {\cal F} \left( \frac{M^2_2}{M^2_3} \right) + {\cal G}  \left( \frac{M^2_2}{M^2_3} \right) \right] \Delta^{}_x \;,
\label{2.6}
\end{eqnarray}
with
\begin{eqnarray}
&& \Delta^{}_x =  c^{}_{23} c^{}_{13} \left[ c^{}_{12} s^{}_{23} \sin \sigma + s^{}_{12} c^{}_{23} s^{}_{13} \sin (\sigma +\delta)  \right] \; ,
\label{2.7}
\end{eqnarray}
which involves only two low-energy CP phases (i.e., $\sigma$ and $\delta$).
Typically, for larger right-handed neutrino masses, the magnitudes of the CP asymmetries increase
[e.g., $\varepsilon^{}_{2 \tau}$ is proportional to $M^{}_3$ as shown in Eq.~(\ref{2.6})]. Given that in the scenario considered in this paper the final baryon asymmetry is suppressed by $y^2_\tau$ as shown in Eq.~(\ref{9}), larger right-handed neutrino masses are needed to increase the magnitudes of the CP asymmetries so that the observed value of $Y^{}_{\rm B}$ can be reached. Considering that in the case under consideration (i.e., $O=O^{}_x$) $N^{}_1$ decouples from leptogenesis, we will pay our attention to the allowed values of the next-to-lightest right-handed neutrino mass $M^{}_2$. In Figure~1(a) and (b) (for the NO and IO cases, respectively), in terms of the red lines, we have shown the minimal values of $M^{}_2$ needed to reproduce the observed value of $Y^{}_{\rm B}$ as functions of the lightest neutrino mass ($m^{}_1$ and $m^{}_3$ in the NO and IO cases, respectively). These results are obtained by allowing the free parameters $\delta$, $\sigma$ and $x$ to vary in the range $[0, 2\pi]$, and the mass ratio $M^{}_3/M^{}_2$ to vary in the range $[3, 10]$. We see that in order to reproduce the observed value of $Y^{}_{\rm B}$ one needs to have $M^{}_2 \gtrsim 10^{14}$ GeV. In Figure~2(a), we have further shown the results in the case that $\delta$ is the only origin of CP violation (with $\sigma=0$). One can see that in this case the right-handed neutrino mass scale should be further lifted (to be around $10^{15}$ GeV) in order to reproduce the observed value of $Y^{}_{\rm B}$. This result can be easily understood with the help of Eq.~(\ref{2.7}): the first term becomes vanishing for $\sigma=0$, while the second term is suppressed by $s^{}_{13}$.

It should be pointed out that although the right-handed neutrino masses are above $10^{14}$ GeV, the leptogenesis calculations still can be performed in the two-flavor regime. This is viable provided that the critique temperature $T^{}_{\rm c}$ at which the right-handed neutrinos acquire their masses are below $10^{12}$ GeV (note that it is $T^{}_{\rm c}$ rather than the right-handed neutrino masses that measures the temperature of the plasma in the Universe and determines which lepton flavor has come into play). If $T^{}_{\rm c}$ were above $10^{12}$ GeV, one would have to work in the unflavored regime and simply get $Y^{}_{\rm B} =0$. Of course, a non-vanishing $Y^{}_{\rm B}$ is still possible when the quantum density matrix formalism is taken into account \cite{matrix}. The latter possibility is beyond the scope of this work, and here we just take $T^{}_{\rm c}$ to be below $10^{12}$ GeV so that one can work in the usual two-flavor regime.

\begin{figure*}
\centering
\includegraphics[width=6.5in]{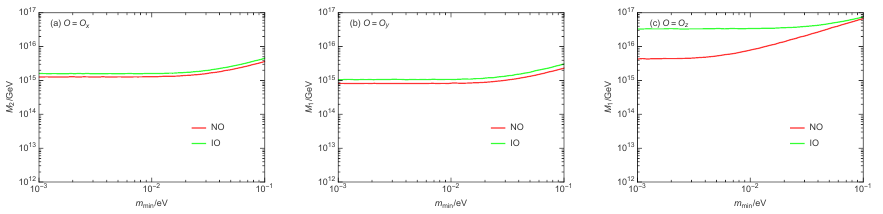}
\caption{ For the cases of $O=O^{}_x$, $O^{}_y$ and $O^{}_z$, the minimal values of the right-handed neutrino masses needed to reproduce the observed value of $Y^{}_{\rm B}$ as functions of the lightest neutrino mass (with $m^{}_{\rm min}= m^{}_1$ or $m^{}_3$ in the NO or IO case) in the case that $\delta$ is the only origin of CP violation. }
\label{fig2}
\end{figure*}

We then perform the study for the case of $O = O^\prime_x$. In this case, one has
\begin{eqnarray}
&& \varepsilon^{}_{2 \tau} = - \frac{M^{}_3 \sqrt{m^{}_2 m^{}_3 } (m^{}_2 + m^{}_3) \sinh 2x  }{16\pi v^2 (m^{}_2 \cosh^2 x+ m^{}_3 \sinh^2 x) }  \left[ {\cal F} \left( \frac{M^2_3}{M^2_2} \right) - {\cal G}  \left( \frac{M^2_3}{M^2_2} \right) \right] \Delta^{\prime}_x \; , \nonumber \\
&& \varepsilon^{}_{3 \tau} = \frac{M^{}_2 \sqrt{m^{}_2 m^{}_3 } (m^{}_2 + m^{}_3) \sinh 2x  }{16\pi v^2 (m^{}_2 \sinh^2 x+ m^{}_3 \cosh^2 x) }  \left[ {\cal F} \left( \frac{M^2_2}{M^2_3} \right) - {\cal G}  \left( \frac{M^2_2}{M^2_3} \right) \right] \Delta^{\prime}_x \;.
\label{2.8}
\end{eqnarray}
with
\begin{eqnarray}
&& \Delta^{\prime}_x =  c^{}_{23} c^{}_{13}  \left[ c^{}_{12} s^{}_{23} \cos \sigma + s^{}_{12} c^{}_{23} s^{}_{13} \cos (\sigma +\delta)  \right] \; .
\label{2.9}
\end{eqnarray}
It is interesting to note that in this case $\varepsilon^{}_{I \tau}$ can be non-vanishing (in fact, take their maximal values) even for $\sigma =0$ and $\delta=0$. Therefore, for this case (and similarly for the cases of $O=O^\prime_y$ and $O^\prime_z$) we will not consider the possibility that $\delta$ is the only origin of CP violation. In Figure~1(a) and (b) (for the NO and IO cases, respectively), in terms of the green lines, we have shown the minimal values of $M^{}_2$ needed to reproduce the observed value of $Y^{}_{\rm B}$ as functions of the lightest neutrino mass. These results are obtained in the same way as in the case of $O=O^{}_x$ except that here we allow $x$ to vary in the range $[-3, 3]$ for the following consideration: large values of $x$ imply a strong fine tuning because they imply that neutrino masses are much lighter than the individual terms $(M^{}_{\rm D})^{2}_{\alpha I}/M^{}_I$ because of sign cancelations. Therefore, such choices tend to transfer the
explanation of neutrino lightness from the seesaw mechanism to some other mechanism
that has to explain the fine-tuned cancelations. A point of view held by Ref.~\cite{BB} is to consider the $O$ matrices to be "reasonable" if $|\sinh x|, |\sinh y|, |\sinh z| \lesssim 1$ (corresponding to $|x|, |y|, |z| \lesssim 1$), and "acceptable" if $|\sinh x|, |\sinh y|, |\sinh z| \lesssim 10$ (corresponding to $|x|, |y|, |z| \lesssim 3$). One can see that the results in the present case are similar to (just a litter larger than) those in the case of $O=O^{}_x$.

In the case of $O=O^{}_y$, $N^{}_2$ will decouple from leptogenesis and its only role is to be responsible for the generation of $m^{}_2$. In this case, one just needs to consider the contributions of $N^{}_1$ and $N^{}_3$ to the final baryon asymmetry, and has
\begin{eqnarray}
&& \varepsilon^{}_{1 \tau} = \frac{M^{}_3 \sqrt{m^{}_1 m^{}_3 } (m^{}_1 - m^{}_3) \sin 2y  }{16\pi v^2 (m^{}_1 \cos^2 y+ m^{}_3 \sin^2 y) }  \left[ {\cal F} \left( \frac{M^2_3}{M^2_1} \right) + {\cal G}  \left( \frac{M^2_3}{M^2_1} \right) \right] \Delta^{}_y \; , \nonumber \\
&& \varepsilon^{}_{3 \tau} = \frac{M^{}_1 \sqrt{m^{}_1 m^{}_3 } (m^{}_3 - m^{}_1) \sin 2y }{16\pi v^2 (m^{}_1 \sin^2 y+ m^{}_3 \cos^2 y) }  \left[ {\cal F} \left( \frac{M^2_1}{M^2_3} \right) + {\cal G}  \left( \frac{M^2_1}{M^2_3} \right) \right] \Delta^{}_y \;,
\label{2.10}
\end{eqnarray}
with
\begin{eqnarray}
&& \Delta^{}_y =  c^{}_{23} c^{}_{13} \left[ c^{}_{12} c^{}_{23} s^{}_{13} \sin (\delta + \rho)
- s^{}_{12} s^{}_{23}  \sin \rho  \right] \;,
\label{2.11}
\end{eqnarray}
which involves $\rho$ and $\delta$.
In the case of $O=O^\prime_y$, one has
\begin{eqnarray}
&& \varepsilon^{}_{1 \tau} = - \frac{M^{}_3 \sqrt{m^{}_1 m^{}_3 } (m^{}_1 + m^{}_3) \sinh 2y  }{16\pi v^2 (m^{}_1 \cosh^2 y+ m^{}_3 \sinh^2 y) }  \left[ {\cal F} \left( \frac{M^2_3}{M^2_1} \right) - {\cal G}  \left( \frac{M^2_3}{M^2_1} \right) \right] \Delta^{\prime}_y \; , \nonumber \\
&& \varepsilon^{}_{3 \tau} = \frac{M^{}_1 \sqrt{m^{}_1 m^{}_3 } (m^{}_1 + m^{}_3) c^{}_{23} c^{}_{13} \sinh 2y }{16\pi v^2 (m^{}_1 \sinh^2 y+ m^{}_3 \cosh^2 y) }  \left[ {\cal F} \left( \frac{M^2_1}{M^2_3} \right) - {\cal G}  \left( \frac{M^2_1}{M^2_3} \right) \right] \Delta^{\prime}_y \;,
\label{2.12}
\end{eqnarray}
with
\begin{eqnarray}
&& \Delta^{\prime}_y =  c^{}_{23} c^{}_{13} \left[ c^{}_{12} c^{}_{23} s^{}_{13} \cos (\delta + \rho)
- s^{}_{12} s^{}_{23}  \cos \rho  \right] \; .
\label{2.13}
\end{eqnarray}
For these two cases, in Figure~1(c) and (d) (for the NO and IO cases, respectively) we have shown the minimal values of $M^{}_1$ needed to reproduce the observed value of $Y^{}_{\rm B}$ as functions of the lightest neutrino mass. And for the case of $O=O^{}_y$, in Figure~2(b) we have further shown the results in the case that $\delta$ is the only origin of CP violation (with $\rho=0$). One can see that the results in these two cases are similar to those in the cases of $O=O^{}_x$ and $O^\prime_x$.

In the case of $O = O^{}_z$, $N^{}_3$ will decouple from leptogenesis and its only role is to be responsible for the generation of $m^{}_3$. In this case, one just needs to consider the contributions of $N^{}_1$ and $N^{}_2$ to the final baryon asymmetry, and has
\begin{eqnarray}
&& \varepsilon^{}_{1 \tau} = \frac{M^{}_2 \sqrt{m^{}_1 m^{}_2 } (m^{}_2 - m^{}_1) \sin 2z  }{16\pi v^2 (m^{}_1 \cos^2 z+ m^{}_2 \sin^2 z) }  \left[ {\cal F} \left( \frac{M^2_2}{M^2_1} \right) + {\cal G}  \left( \frac{M^2_2}{M^2_1} \right) \right] \Delta^{}_z \; , \nonumber \\
&& \varepsilon^{}_{2 \tau} = \frac{M^{}_1 \sqrt{m^{}_1 m^{}_2 } (m^{}_1 - m^{}_2) \sin 2z  }{16\pi v^2 (m^{}_1 \sin^2 z+ m^{}_2 \cos^2 z) }  \left[ {\cal F} \left( \frac{M^2_1}{M^2_2} \right) + {\cal G}  \left( \frac{M^2_1}{M^2_2} \right) \right] \Delta^{}_z \;,
\label{2.14}
\end{eqnarray}
with
\begin{eqnarray}
&& \Delta^{}_z =    c^{}_{23} s^{}_{23} s^{}_{13} [ c^2_{12}\sin (\delta + \rho -\sigma)
+ s^2_{12} \sin (\delta - \rho + \sigma) ]
+ c^{}_{12} s^{}_{12} ( c^{2}_{23}  s^{2}_{13} - s^{2}_{23} ) \sin (\rho -\sigma) \;,
\label{2.15}
\end{eqnarray}
which involves $\rho-\sigma$ and $\delta$.
In the case of $O = O^\prime_z$, one has
\begin{eqnarray}
&& \varepsilon^{}_{1 \tau} = \frac{M^{}_2 \sqrt{m^{}_1 m^{}_2 } (m^{}_1 + m^{}_2) \sinh 2z  }{16\pi v^2 (m^{}_1 \cosh^2 z+ m^{}_2 \sinh^2 z) }  \left[ {\cal F} \left( \frac{M^2_2}{M^2_1} \right) - {\cal G}  \left( \frac{M^2_2}{M^2_1} \right) \right] \Delta^{\prime}_z \; , \nonumber \\
&& \varepsilon^{}_{2 \tau} = - \frac{M^{}_1 \sqrt{m^{}_1 m^{}_2 } (m^{}_1 + m^{}_2) \sinh 2z  }{16\pi v^2 (m^{}_1 \sinh^2 z+ m^{}_2 \cosh^2 z) }  \left[ {\cal F} \left( \frac{M^2_1}{M^2_2} \right) - {\cal G}  \left( \frac{M^2_1}{M^2_2} \right) \right] \Delta^{\prime}_z  \;,
\label{2.16}
\end{eqnarray}
with
\begin{eqnarray}
&& \Delta^{\prime}_z =   c^{}_{23} s^{}_{23} s^{}_{13} [ c^2_{12}\cos (\delta + \rho -\sigma)
- s^2_{12} \cos (\delta - \rho + \sigma) ]
+ c^{}_{12} s^{}_{12} ( c^{2}_{23}  s^{2}_{13} - s^{2}_{23} ) \cos (\rho -\sigma) \; .
\label{2.17}
\end{eqnarray}
For these two cases, in Figure~1(e) and (f) (for the NO and IO cases, respectively) we have shown the minimal values of $M^{}_1$ needed to reproduce the observed value of $Y^{}_{\rm B}$ as functions of the lightest neutrino mass. And for the case of $O=O^{}_z$, in Figure~2(c) we have further shown the results in the case that $\delta$ is the only origin of CP violation (with $\rho-\sigma=0$).
One can see that the results in these two cases are larger than the corresponding results in the cases of $O=O^{}_x, O^\prime_x, O^{}_y$ and $O^\prime_y$ by about one order of magnitude. These results can be easily understood with the help of Eqs.~(\ref{2.15}, \ref{2.17}): the first term of $\Delta^{}_z$ and $\Delta^\prime_z$ is suppressed by $s^{}_{13}$, while the second term of them is also subject to a suppression that is due to the accidental cancelation between $c^{2}_{23}  s^{2}_{13}$ and  $s^{2}_{23}$ (for $\theta^{}_{23} \sim \pi/4$ and small $\theta^{}_{13}$).

\section{Study for nearly degenerate right-handed neutrino masses}

In this section, we perform the study in the scenario that the right-handed neutrino masses are nearly degenerate. In this scenario, the flavored CP asymmetries will get resonantly enhanced as \cite{resonant}
\begin{eqnarray}
\varepsilon^{}_{I\alpha} = \frac{{\rm Im}\left\{ (M^*_{\rm D})^{}_{\alpha I} (M^{}_{\rm D})^{}_{\alpha J}
\left[ M^{}_J (M^\dagger_{\rm D} M^{}_{\rm D})^{}_{IJ} + M^{}_I (M^\dagger_{\rm D} M^{}_{\rm D})^{}_{JI} \right] \right\} }{8\pi  v^2 (M^\dagger_{\rm D} M^{}_{\rm D})^{}_{II}} \cdot \frac{M^{}_I \Delta M^2_{IJ}}{(\Delta M^2_{IJ})^2 + M^2_I \Gamma^2_J} \;,
\label{3.1}
\end{eqnarray}
where $\Delta M^2_{IJ} \equiv M^2_I - M^2_J$ has been defined and $\Gamma^{}_J= (M^\dagger_{\rm D} M^{}_{\rm D})^{}_{JJ} M^{}_J/(8\pi v^2)$ is the decay rate of $N^{}_J$ (for $J \neq I$).
Thanks to such a resonance enhancement effect, a successful leptogenesis can be realized even with TeV-scale right-handed neutrinos which have the potential to be directly accessed by running or upcoming experiments \cite{RHN}. Correspondingly, the total CP asymmetries are obtained as
\begin{eqnarray}
\varepsilon^{}_{I} = \frac{{\rm Im}\left\{ (M^\dagger_{\rm D} M^{}_{\rm D})^{}_{I J}
\left[ M^{}_J (M^\dagger_{\rm D} M^{}_{\rm D})^{}_{IJ} + M^{}_I (M^\dagger_{\rm D} M^{}_{\rm D})^{}_{JI} \right] \right\} }{8\pi  v^2 (M^\dagger_{\rm D} M^{}_{\rm D})^{}_{II}} \cdot \frac{M^{}_I \Delta M^2_{IJ}}{(\Delta M^2_{IJ})^2 + M^2_I \Gamma^2_J} \;.
\label{3.2}
\end{eqnarray}
It is also direct to verify that the forms of $O$ in Eq.~(\ref{2.4}) will render such $\varepsilon^{}_{I}$ to be vanishing, realizing the purely flavored leptogenesis scenario.

Let us first perform the study for the case of $O=O^{}_x$. In this case, one has
\begin{eqnarray}
&& \varepsilon^{}_{2\tau} = \frac{ \sqrt{m^{}_2 m^{}_3} (m^{}_3 - m^{}_2) \sin 2 x }{4\pi  v^2 (m^{}_2 \cos^2 x+ m^{}_3 \sin^2 x)} \cdot \frac{M^{2}_0 \Delta M}{4(\Delta M)^2 +  \Gamma^2_3}  \Delta^{}_x \; , \nonumber \\
&& \varepsilon^{}_{3\tau} = \frac{ \sqrt{m^{}_2 m^{}_3} (m^{}_3 - m^{}_2)  \sin 2 x }{4\pi  v^2 (m^{}_2 \sin^2 x+ m^{}_3 \cos^2 x)} \cdot \frac{M^{2}_0 \Delta M }{4(\Delta M)^2 +  \Gamma^2_2}  \Delta^{}_x \; ,
\label{3.3}
\end{eqnarray}
where $M^{}_0 \simeq M^{}_2 \simeq M^{}_3$, $\Delta M= M^{}_3 - M^{}_2$ and $\Delta^{}_x$ has been given in Eq.~(\ref{2.7}). For this case, in Figure~3(a) and (b) (for the NO and IO cases, respectively) we have shown the values of $\Delta M/M^{}_0$ that allow for a successful reproduction of the observed value of $Y^{}_{\rm B}$ as functions of the lightest neutrino mass. These results are obtained by taking $M^{}_0=1$ TeV as a benchmark value and allowing the other related parameters to vary in their respective allowed ranges. One can see that $\Delta M/M^{}_0$ needs to fall into the range $10^{-16}-10^{-12}$ in order to allow for a successful reproduction of the observed value of $Y^{}_{\rm B}$. Here we would like to point out that, although we have taken $M^{}_0=1$ TeV as the benchmark value, the results of $Y^{}_{\rm B}$ will keep invariant for other values of $M^{}_0$ provided that the combination $\Delta M/M^2_0$ takes a constant value. This can be easily understood with the help of Eq.~(\ref{3.3}): in the expressions of $\varepsilon^{}_{I\tau}$, $\Delta M$ and $M^{}_0$ only take effect in the form of $\Delta M/M^2_0$ (note that $\Gamma^{}_I$ is proportional to $M^{2}_I$). In other words, when $M^{}_0$ takes another value, the values of $\Delta M/M^{}_0$ that allow for a successful reproduction of the observed value of $Y^{}_{\rm B}$ can be obtained from those in Figure~3 by means of a simple rescaling law (e.g., will be lifted by 10 times for $M^{}_0=10$ TeV).

In Figure~4(a) and (b) (for the NO and IO cases, respectively), we have further shown the values of $\delta$ that allow for a successful reproduction of the observed value of $Y^{}_{\rm B}$ as functions of the lightest neutrino mass in the case that $\delta$ is the only origin of CP violation (with $\sigma=0$). It is interesting to note that there is an upper bound about 0.07 eV for $m^{}_1$ and $m^{}_3$, and at here $\delta$ needs to be around $\pm \pi/2$ in order to allow for a successful reproduction of the observed value of $Y^{}_{\rm B}$.

\begin{figure*}
\centering
\includegraphics[width=6.5in]{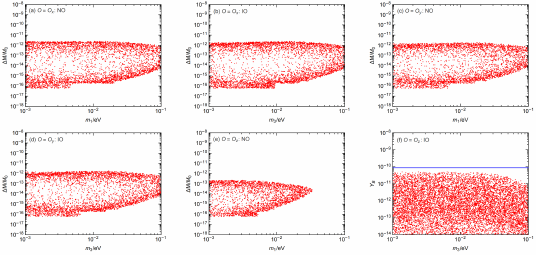}
\caption{ (a)-(e): For the cases of $O=O^{}_x$, $O^{}_y$ and $O^{}_z$, the values of $\Delta M/M^{}_0$ that allow for a successful reproduction of the observed value of $Y^{}_{\rm B}$ as functions of the lightest neutrino mass. (f): For the case of $O=O^{}_z$, the allowed values of $Y^{}_{\rm B}$ as functions of the lightest neutrino mass $m^{}_3$ in the IO case. The blue horizontal line stands for the observed value of $Y^{}_{\rm B}$.
These results are obtained by taking $M^{}_0=1$ TeV as the benchmark value, while the other related parameters are allowed to vary in their respective allowed ranges. }
\label{fig3}
\end{figure*}

\begin{figure*}
\centering
\includegraphics[width=6.5in]{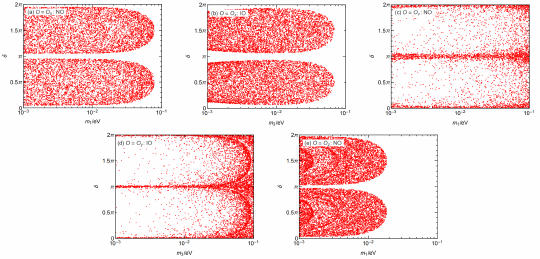}
\caption{ For the cases of $O=O^{}_x$, $O^{}_y$ and $O^{}_z$, the values of $\delta$ that allow for a successful reproduction of the observed value of $Y^{}_{\rm B}$ as functions of the lightest neutrino mass in the case the $\delta$ is the only origin of CP violation. }
\label{fig4}
\end{figure*}

We then perform the study for the case of $O = O^\prime_x$. In this case, one has
\begin{eqnarray}
&& \varepsilon^{}_{2\tau} = \frac{ \sqrt{m^{}_2 m^{}_3} (m^{}_2 + m^{}_3) \sinh 2 x }{8\pi  v^2 (m^{}_2 \cosh^2 x+ m^{}_3 \sinh^2 x)} \cdot \frac{M^{}_0 (\Delta M)^2}{4(\Delta M)^2 +  \Gamma^2_3}  \Delta^{\prime}_x \; , \nonumber \\
&& \varepsilon^{}_{3\tau} = \frac{ \sqrt{m^{}_2 m^{}_3} (m^{}_2 + m^{}_3)  \sinh 2 x }{8\pi  v^2 (m^{}_2 \cosh^2 x+ m^{}_3 \sinh^2 x)} \cdot \frac{M^{}_0 (\Delta M)^2 }{4(\Delta M)^2 +  \Gamma^2_2}  \Delta^{\prime}_x \; ,
\label{3.4}
\end{eqnarray}
where $M^{}_0 \simeq M^{}_2 \simeq M^{}_3$, $\Delta M= M^{}_3 - M^{}_2$ and $\Delta^{\prime}_x$ has been given in Eq.~(\ref{2.9}). A comparison between Eq.~(\ref{3.3}) and Eq.~(\ref{3.4}) shows that $\varepsilon^{}_{I\tau}$ are additionally suppressed by $\Delta M/M^{}_0$ (which is very small as demanded by the resonance condition) in the case of $O = O^\prime_x$ compared to in the case of $O = O^{}_x$. Hence it is natural to expect that the observed value of $Y^{}_{\rm B}$ cannot be reached in the case of $O = O^\prime_x$ (and we have numerically checked this). The results for the cases of $O = O^\prime_y$ and $O^\prime_z$ are similar, and hence we will not present them in the following.

In the case of $O=O^{}_y$, one has
\begin{eqnarray}
&& \varepsilon^{}_{1\tau} = \frac{ \sqrt{m^{}_1 m^{}_3} (m^{}_3 - m^{}_1) \sin 2 y }{4\pi  v^2 (m^{}_1 \cos^2 y+ m^{}_3 \sin^2 y)} \cdot \frac{M^{2}_0 \Delta M}{4(\Delta M)^2 +  \Gamma^2_3}  \Delta^{}_y \; , \nonumber \\
&& \varepsilon^{}_{3\tau} = \frac{ \sqrt{m^{}_1 m^{}_3} (m^{}_3 - m^{}_1)  \sin 2 y }{4\pi  v^2 (m^{}_1 \sin^2 y+ m^{}_3 \cos^2 y)} \cdot \frac{M^{2}_0 \Delta M}{4(\Delta M)^2 +  \Gamma^2_1}  \Delta^{}_y \; ,
\label{3.5}
\end{eqnarray}
where $M^{}_0 \simeq M^{}_1 \simeq M^{}_3$, $\Delta M= M^{}_3 - M^{}_1$ and $\Delta^{}_y$ has been given in Eq.~(\ref{2.11}). For this case, in Figure~3(c) and (d) (for the NO and IO cases, respectively) we have shown the values of $\Delta M/M^{}_0$ that allow for a successful reproduction of the observed value of $Y^{}_{\rm B}$ as functions of the lightest neutrino mass. One can see that in this case $\Delta M/M^{}_0$ also needs to fall into the range $10^{-16}-10^{-12}$ in order to allow for a successful reproduction of the observed value of $Y^{}_{\rm B}$. In Figure~4(c) and (d) (for the NO and IO cases, respectively), we have further shown the values of $\delta$ that allow for a successful reproduction of the observed value of $Y^{}_{\rm B}$ as functions of the lightest neutrino mass in the case that $\delta$ is the only origin of CP violation (with $\rho=0$). It is interesting to note that $\delta$ needs to be around $0$ or $\pi$ in order to allow for a successful reproduction of the observed value of $Y^{}_{\rm B}$.

In the case of $O =O^{}_z$, one has
\begin{eqnarray}
&& \varepsilon^{}_{1\tau} = \frac{ \sqrt{m^{}_1 m^{}_2} (m^{}_1 - m^{}_2) \sin 2 z }{4\pi  v^2 (m^{}_1 \cos^2 z+ m^{}_2 \sin^2 z)} \cdot \frac{M^{2}_0 \Delta M}{4(\Delta M)^2 +  \Gamma^2_2}  \Delta^{}_z \; , \nonumber \\
&& \varepsilon^{}_{2\tau} = \frac{ \sqrt{m^{}_1 m^{}_2} (m^{}_1 - m^{}_2)  \sin 2 z }{4\pi  v^2 (m^{}_1 \sin^2 z+ m^{}_2 \cos^2 z)} \cdot \frac{M^{2}_0 \Delta M}{4(\Delta M)^2 +  \Gamma^2_1}  \Delta^{}_z \; ,
\label{3.6}
\end{eqnarray}
where $M^{}_0 \simeq M^{}_1 \simeq M^{}_2$, $\Delta M= M^{}_2 - M^{}_1$ and $\Delta^{}_z$ has been given in Eq.~(\ref{2.15}).
For this case, in Figure~3(e) we have shown the values of $\Delta M/M^{}_0$ that allow for a successful reproduction of the observed value of $Y^{}_{\rm B}$ as functions of the lightest neutrino mass $m^{}_1$ in the NO case. One can see that in this case the parameter space of leptogenesis shrinks: in order to allow for a successful reproduction of the observed value of $Y^{}_{\rm B}$, $\Delta M/M^{}_0$ needs to fall into the range $10^{-16}-10^{-13}$ while $m^{}_1$ should be smaller than 0.03 eV. Such a shrink is also due to the suppression of $\Delta^{}_z$ (the first term by $s^{}_{13}$ and the second term by the accidental cancelation between $c^{2}_{23}  s^{2}_{13}$ and $s^{2}_{23}$). In Figure~4(e) we have further shown the values of $\delta$ that allow for a successful reproduction of the observed value of $Y^{}_{\rm B}$ as functions of $m^{}_1$ in the case that $\delta$ is the only origin of CP violation (with $\rho-\sigma=0$). One can see that there is an upper bound about 0.02 eV for $m^{}_1$, and at here $\delta$ needs to be around $\pm \pi/2$ in order to allow for a successful reproduction of the observed value of $Y^{}_{\rm B}$. On the other hand, in the IO case the observed value of $Y^{}_{\rm B}$ cannot be reached, as shown by Figure~3(f). This result can be easily understood with the help of Eq.~(\ref{3.6}): in the IO case, $\varepsilon^{}_{I\tau}$ are subject to a further suppression from $m^{}_1 - m^{}_2$ (as a result of the approximate equality between $m^{}_1$ and $m^{}_2$).

\section{Summary}

In the literature, a physically interesting and extensively studied possibility is when the leptogenesis mechanism is completely realized through the so-called flavor effects (i.e., purely flavored leptogenesis). In this scenario, although the flavored CP asymmetries $\varepsilon^{}_{I\alpha}$ are non-vanishing individually, their flavored sums $\varepsilon^{}_{I}$ are vanishing (which can
be naturally realized in flavour models with residual CP symmetries). As a result, the successful leptogenesis cannot be realized in the unflavored regime, but may be achieved in the two-flavor or three-flavor regime thanks to the flavor non-universality of the washout effects.

Another physically interesting possibility is that the right-handed neutrino are initially protected to be massless by some symmetry and then suddenly become massive at a critique temperature. If the gained masses of the right-handed neutrinos are much larger than the temperature of the Universe at that time, they will decay and generate a nonzero lepton asymmetry very rapidly. And in this scenario the washout effects for the generated lepton asymmetry can be neglected safely.

In this paper, we have pointed out that in the scenario that the right-handed neutrinos suddenly gain some masses much larger than the temperature of the Universe at that time so that the washout effects for the generated lepton asymmetry can be  neglected safely, the purely flavored leptogenesis (which crucially relies on the flavor non-universality of the washout effects) cannot work in the usual way any more. For this problem, we put forward that the flavor non-universality of the conversion efficiencies [see Eq.~(\ref{8})] from the lepton asymmetries to the baryon asymmetry via the sphaleron processes may play a crucial role.
And we have studied if the requisite baryon asymmetry can be successfully reproduced from such a mechanism in the scenarios that the right-handed neutrino masses are hierarchical and nearly degenerate, respectively.

In the scenario that the right-handed neutrino masses are hierarchical, it is found that the right-handed neutrino masses should be above $10^{14}$ GeV ($10^{15}$ GeV) in the cases of $O=O^{}_x, O^\prime_x, O^{}_y$ and $O^\prime_y$ ($O^{}_z$ and $O^\prime_z$) in order to reproduce the observed value of $Y^{}_{\rm B}$. And in the case that $\delta$ is the only origin of CP violation, the right-handed neutrino mass scale should be further lifted in order to reproduce the observed value of $Y^{}_{\rm B}$.

In the scenario that the right-handed neutrino masses are nearly degenerate, for the cases of $O=O^{}_x$, $O^{}_y$ and $O^{}_z$ (only in the NO case), the successful leptogenesis can be achieved for TeV-scale right-handed neutrinos with mass degeneracies in the range $10^{-16}-10^{-12}$. And the particular case that $\delta$ is the only origin of CP violation can be viable. On the other hand, for the cases of $O=O^{\prime}_x$, $O^{\prime}_y$, $O^{\prime}_z$ and $O^{}_z$ (only in the IO case), the observed value of $Y^{}_{\rm B}$ cannot be reached.

\vspace{0.5cm}

\underline{Acknowledgments} \vspace{0.2cm}

This work is supported in part by the National Natural Science Foundation of China under grant NO.~11605081, the Natural Science Foundation of the Liaoning Scientific Committee under grant NO.~2022-MS-314, and the Basic Research Business Fees for Universities in Liaoning Province (2024).

\end{document}